\documentclass[
 twocolumn
]{ceurart}

\usepackage{listings}
\usepackage{xcolor}

\usepackage{soul}
\usepackage{mathtools}
\usepackage{hyperref}
\usepackage{amsmath}

\begin{document}

%%
%% Rights management information.
%% CC-BY is default license.
\copyrightyear{2024}
\copyrightclause{Copyright for this paper by its authors.
  Use permitted under Creative Commons License Attribution 4.0
  International (CC BY 4.0).}

%%
%% This command is for the conference information
\conference{IMSC Workshop@JCDL 2024,
 December 20, 2024, Hong Kong, CN}

%%
%% The "title" command
\title{Freshness and Informativity Weighted Cognitive Extent and Its Correlation with Cumulative Citation Count}

%%
%% The "author" command and its associated commands are used to define
%% the authors and their affiliations.
\author[1]{Zihe Wang}[%
orcid=0009-0007-4293-3449,
email=wang.14629@osu.edu,
]
\fnmark[1]
\address[1]{Ohio State University, Columbus, OH, United States}

\author[2]{Jian Wu}[%
orcid=0000-0003-0173-4463,
email=j1wu@odu.edu,
url=https://www.cs.odu.edu/~jwu/,
]
\cormark[1]
\fnmark[1]
\address[2]{Old Dominion University, Norfolk, VA, United States}

%% Footnotes
\cortext[1]{Corresponding author.}
\fntext[1]{These authors contributed equally.}

%%
%% The abstract is a short summary of the work to be presented in the
%% article.
\begin{abstract}
In this paper, we revisit cognitive extent, originally defined as the number of unique phrases in a quota. We introduce Freshness and Informative Weighted Cognitive Extent (FICE), calculated based on two novel weighting factors, the lifetime ratio and informativity of scientific entities. We model the lifetime of each scientific entity as the time-dependent document frequency, which is fit by the composition of multiple Gaussian profiles. The lifetime ratio is then calculated as the cumulative document frequency at the publication time $t_0$ divided by the cumulative document frequency over its entire lifetime. The informativity is calculated by normalizing the document frequency across all scientific entities recognized in a title. Using the ACL Anthology, we verified the trend formerly observed in several other domains that the number of unique scientific entities per quota increased gradually at a slower rate. We found that FICE exhibits a strong correlation with the average cumulative citation count within a quota. Our code is available at \href{https://github.com/ZiheHerzWang/Freshness-and-Informativity-Weighted-Cognitive-Extent}{https://github.com/ZiheHerzWang/Freshness-and-Informativity-Weighted-Cognitive-Extent}
\end{abstract}

%%
%% Keywords. The author(s) should pick words that accurately describe
%% the work being presented. Separate the keywords with commas.

%%
%% This command processes the author and affiliation and title
%% information and builds the first part of the formatted document.
\maketitle

\section{Introduction}
Cognitive extent is an approach to quantify the extent of cognitive domains of scientific fields based on the concept of lexical diversity \cite{milojevic2015quantifying}. 
The metric was originally calculated by counting the number of unique concepts (phrases) appearing in the titles of statistically large unit quotas of scientific literature, which reflects the extent of the cognitive territory covered in that literature. 
%%A phrase is defined as the longest string of words separated by general words or a phrase delimiter, such as punctuation marks. 
Cognitive extent has been used as a representation of knowledge gained by scientists. It has been shown that cognitive extent calculated in multiple academic fields (Physics, Astronomy, and Biomedicine) grew at a slower rate.
%%This metric overcame the limitation of an early metric which determined the lexical diversity only by single words because it captures new concepts from the combination of existing words. \cite{jarvis2013defining, malvern2004lexical}. 
%%, which is in agreement with the notation made by Price  (1963) and the recent emphasis that ``ideas getting harder to find'' \cite{bloom2020ideas}. 

However, this definition of cognitive extent has two limitations. First, it only accounts for the occurrence of a phrase as a dichotomous value within a quota and ignores \emph{when} the phrase occurs. Specifically, a phrase is novel when it appears the first time in the title of Paper A, but when it appears again in the title of Paper B, it still maintains a level of freshness, and thus still reflects the scientist of Paper B's cognitive knowledge except that the knowledge is no longer new. Second, the definition treats all phrases with the same weight. However, certain phrases may be more informative than others. For example, at a particular time the phrase ``entity recognition'' occurred in many titles but the phrase ``nuclearity rhetorical relation'' occurred in only a small number of titles. From a reader's perspective, the latter phrase is more informative because the former phrase has been seen in many papers.

%%However, a title usually contains a limited number of concepts and the calculation cognitive extent only counts unique concepts, which is equivalent to setting a binary label (0 or 1) for each concept and only counting the concepts labeled as 1. Using this method, if a title contains concepts that already appeared in previous papers, it does not have any contribution to the ``knowledge'' growth. However, scientific papers may combine concepts and make contributions in terms of methodology, which is commonly seen in artificial intelligence (AI) papers. On the other hand, authors may coin new concepts that are never used in the titles of new papers, making them look unique but not contributing to knowledge growth. 

To overcome the limitations, we propose Freshness and Informativity Weighted Cognitive Extend (FICE), which is calculated as a weighted occurrence of disambiguated unique scientific entities extracted from paper titles in a quota. Different from the original cognitive extent, FICE accounts for contributions of \emph{freshness} and \emph{informativity} of scientific entities extracted from documents within a corpus. The freshness is based on the lifetime ratio and the informativity is based on time-dependent document frequency across scientific entities in a document. Here the document can be any form of scholarly text. Throughout this paper, we focus our study on paper titles. 

We verified a previous finding using ACL Anthology papers that although the number of papers increased exponentially, the unique number of scientific entities per quota gradually increased with time at a slower rate. One property of FICE is its relationship with citation impact factors. We found that FICE exhibits a strong correlation with the logarithm of 5-year average cumulative citation counts for papers in the ACL Anthology.

%%Our contributions are summarized below. 
%%\begin{enumerate}
%%    \item We proposed Freshness and Informativity Weighted Cogntive Extent (FICE) that incorporates the freshness and informativity of scientific entities, to quantify the extent of knowledge of scientific fields within a quota and studied its properties using papers in the ACL Anthology. 
%%    \item We found that FICE exhibits a strong correlation with the logarithm of 5-year average cumulative citation counts for papers in the ACL Anthology.
%%    \item We verified that although the number of papers increased exponentially from 1980 to 2020, the unique number of scientific entities per quota gradually increased with time at a lower rate, which is consistent with previous observations in several other domains.
%%\end{enumerate}

\section{Related Work}
The definition of cognitive extent is closely relevant to lexical diversity, defined as the extent of vocabulary disparity within a given language sample \cite{berube2018}. Early metrics determined the lexical diversity only by single words\cite{jarvis2013defining, malvern2004lexical}. Berube et al. proposed the type-token ratio as a metric of lexical diversity \cite{berube2018}. Both methods were based on word-level tokens and ignored their connections, therefore they do not necessarily reflect the \emph{knowledge}, which is better captured by phrases and entities. 

Milojevi\'{c} was the first to propose using concepts (phrases) to quantify the extent of cognitive domains of scientific fields \cite{milojevic2015quantifying}. This method was later applied to study the properties of paper titles in various domains \cite{milojevic2017length}. Recently, a method was proposed to use neural embedding of paper titles to represent the cognitive content of a cluster of papers \cite{eykens2023cognitive}. Although neural embedding has been widely used to capture the semantics of text and compare the semantic similarities between two pieces of text, the embedding itself cannot be directly converted to a numeral that represents the cognitive extent of a single or a corpus of documents. 

Bibliographic impact factors have been extensively studied and several citation impact factors have been proposed \cite{waltman2016review}. The total number of citations (raw citations) is usually criticized as a good indicator because of domain discrepancies, data completeness, and other random factors. The average number of citations per publication of a research unit is frequently used but also criticized because the average value can be biased due to the skewness of citation distribution within the research unit \cite{albarran2011skewness}. However, it was argued that short-term citations can be considered as currency on the research front and long-term citations can contribute to the codification of knowledge claims into bodies of knowledge \cite{leydesdorff2016citations}. Determining the exact boundary between short-term and long-term is non-trivial. Here we adopt $C_5 (y)$, which is the 5-year average cumulative citation count of papers within a quota as an estimate of the quota's average impact over the short- and long-term.

%%The number of highly cited publications is usually considered an alternative to the average because it measures the level of scientific excellence \cite{bormann2014how}. Other citation impact indicators include the proportion of highly cited publications, the h-index \cite{hirsch2005hindex}, and its variants, e.g., g-index \cite{egghe2006gindex}. In our paper, we use  $C_5$, which is the 5-year cumulative average cumulative citation count of papers within a quota as an estimate of the overall impact of papers within the quota. 
%%title conventions in title design: https://doi.org/10.18552/joaw.v5i1.168

\section{Methodology}
%%FICE is calculated based on the lifetime ratio and the informativity weight of scientific entities.

\subsection{Scientific Entity Recognition and Disambiguation}\label{sec:ner}
%%In the original cognitive extent, a phrase in the title was defined as the longest string of words separated by general word(s), or a phrase delimiter such as punctuation marks \cite{milojevic2015quantifying}. This definition, although conceptually clear, is non-trivial to be implemented. Specifically, a list of general words and common words needs to be manually constructed based on the statistics within a corpus. These lists can be domain-dependent. In addition, word sense disambiguation was not incorporated in the extraction process. Therefore, phrases such as ``deep learning'', and ``deep neural networks'' are treated as distinct phrases although they have very close semantic meanings.

%%Therefore, to calculate FICE, we use disambiguated \emph{scientific entities} from a given document. 
A scientific entity is defined as a noun phrase that delivers domain knowledge of interest \cite{wu2020comparative,hong2020sciner}. Scientific entities can be extracted using sequence tagging models, and constructing knowledge graphs, e.g., \cite{li2020nersurvey,hong2020sciner}. Recently, large language models (LLMs) have shown superior performance on named entity recognition, e.g., \cite{wang2023gptner,sousa2023ner}. 

Two scientific entities may have similar semantics. To investigate the impact of entity disambiguation on the results, we conflate entities with similar semantic meanings. This was treated as a classification problem using a thresholded method on the similarity scores calculated using the Cross Encoder \cite{li2020deep}, a model that takes two entity names and outputs a similarity score. The threshold was calibrated based on the classification performance evaluated on a set of manually labeled entity pairs.

%%Scientific entities can be disambiguated using language models, e.g., SciBERT \cite{beltagy2019scibert}. Specifically, the language model can convert scientific entities into dense vectors. The similarities of two scientific entities can be calculated using cosine similarity between their vectors. Unique scientific entities are then identified by conflating entity pairs whose semantic similarities are above a threshold.

\subsection{Lifetime Ratio}
The lifetime of a scientific entity is defined as the period during which it appears in at least one document (in our case, a paper title). Here, we assume that all scientific entities have a finite lifetime, meaning there is a time point $t_s$ when a scientific entity first appears and another time point $t_e$ after which it no longer presents in any documents. We borrowed the concept of document frequency from information retrieval to indicate whether a scientific entity $e$ is still within its lifetime ($df(e,t)>0$) or its lifetime ends ($df(e,t)\leq0$). Here, we define the time-dependent document frequency $df(e,t)$, which is the number of documents that contain the scientific entity $e$ and are published at time $t$ (e.g., a certain year). The lifetime ratio of a scientific entity $e$ at $t_0$ is then defined as the number of accumulated documents up to  $t_0$ divided by the total number of documents that contain $e$ over its entire lifetime $[t_s, t_e]$, 
\begin{equation}
    r(e,t_0)=\frac{\sum_{t_s}^{t_0}df(e,t)}{\sum_{t_s}^{t_e}df(e,t)}.
    \label{eq:r}
\end{equation}
The freshness of $e$ is then calculated as $1-r(e,t_0)$. Because any corpus can cover a limited period, the lifetime ratio can only be calculated based on the \emph{observable period} covered by the corpus. We then model $df(e,t)$ for each $e$ as a composite of analytical profiles. By definition, the lifetime ratio provides an estimation of the relative freshness of a scientific entity. Specifically, a low lifetime ratio indicates a scientific entity is relatively new (so $1-r(e,t_0)$ is large) and a high lifetime ratio indicates a scientific entity is not new anymore. The model fitting provides a prediction of $df(e,t)$ beyond the observable period based on data in the observable period. 

\subsection{Informativity Weight}
In linguistics, informativity concerns the extent to which the contents of a text are already known or expected as compared to unknown or unexpected \cite{de1981introduction}. The informativity of paper titles has been commonly calculated by counting the number of ``substantive'' words. A diachronic analysis of informativity was conducted on chemical paper titles \cite{tocatlian1970titles}, in which non-substantive words were defined as words that convey little or no information, such as articles, prepositions, conjunctions, pronouns, and auxiliary verbs. This objective approach was then extensively used by scholars to study title informativity. We argue that whether a text (word, phrase, entity name) is informative or not depends not only on its semantics but also on the relative frequency it appears in existing papers. At a certain time point $t$, a scientific entity that appears in a large number of documents conveys relatively less new information than an entity that appears only in a few documents. We calculate informativity as the cumulative time-dependent document frequency $DF$ normalized by its range of $[DF_{\rm min}, DF_{\rm max}]$ across all scientific entities in a document,
\begin{equation}
\begin{split}
    w(e,t_0) &= 1-\frac{DF-DF_{\rm min}}{DF_{\rm max}-DF_{\rm min}},\quad
    DF(e,t_0)=\sum^{t_0}_{t_s}df(e,t)\\
    DF_{\rm min}&=\min{\{DF(e_i,t_0), e_i\in E\}}\\
    DF_{\rm max}&=\max{\{DF(e_i,t_0), e_i\in E\}}
    \label{eq:w}
\end{split}
\end{equation}
in which $E$ is the set of scientific entities extracted from a document. The FICE of documents in a quota $Q$ is calculated as 
\begin{equation}
    FICE=\sum_{d\in Q}\sum_{e\in d}w(e,t_d)\left(1-r(e,t_d)\right),
    \label{eq:cce}
\end{equation}
in which $t_d$ is the time when document $d$ was published. 

\section{Data Processing}
\subsection{Data Collection}
We downloaded the metadata of all papers in the ACL Anthology \cite{radev2013acl} in {BibTeX} format\footnote{\url{https://aclanthology.org/anthology+abstracts.bib.gz}}. The publication year and title of each paper were extracted from the BibTeX file. The number of papers published each year from 1952 to 2020 is shown in Figure~\ref{fig:paperentityyear}. 
%%Proceeding front matters are mixed with research papers but only take a fraction of $<1\%$ based on a regular expression matching, which is small enough and is not anticipated to significantly impact the results.
%%It is noteworthy that though the dataset contains abstracts for most of the papers and it concludes the paper better, there are still many of them are not universally available. Consequently, the decision was made to focus on paper titles, which are consistently available and provide a succinct summary of each paper's content.

%%\subsection{Data Processing Pipeline}
%%The modules used to build the data processing pipeline are elaborated below.
%%scientific entity recognition and disambiguation, document frequency curve fitting, and FICE calculation. We also calculated $C_5$ for papers published in Year 2015.
%%, the process begins with data pre-processing in consist of collecting and cleaning the dataset. This is followed by area ratio calculation and concludes to the final novelty score calculation. Each of these stages is crucial for transforming raw data into meaningful metrics. The details will be elaborated in the subsequent sections.

%\begin{figure}[h]
%%    \centering
%%    \includegraphics[width=\linewidth]{pipeline.png}
%%    \caption{Pipeline for Novelty Score Calculation}
%%    \label{fig:pipeline}
%%\end{figure}

%%\begin{figure}
%%    \centering
%%    \includegraphics[width=0.5\textwidth]{figs/prompt.pdf}
%%    \caption{The prompt template used for extracting scientific entities from titles using the GPT4 API.}
%%    \label{fig:prompt}
%%\end{figure}

\subsection{Scientific Entity Recognition}
We compared three off-the-shelf models for scientific entity recognition. (1) \textbf{GPT-4} \cite{brown2020language}.  We construct a zero-shot prompting template to extract scientific entities using GPT-4. The temperature is set to zero to ensure consistent outputs. (2) \textbf{SciBERT} \cite{beltagy2019scibert}. We used the named entity recognition implementation from the Hugging Face library developed by AllenAI\footnote{\url{https://huggingface.co/allenai/scibert_scivocab_cased}}. (3) \textbf{SpaCy} \cite{spacyner}. Entity recognition is performed by invoking the entity recognition module from the Hugging Face library\footnote{\url{https://huggingface.co/spacy/en_core_web_sm}}. 
%%\begin{itemize}
%%    \item , which is based on a transformer model trained on massive amounts of text from the Web, is so far among the state-of-the-art LLMs on many tasks \cite{guo2023evaluating}. We construct prompts to GPT-4 to extract scientific entities from titles. A prompting template is shown in Figure~\ref{fig:prompt}. The temperature is set to zero to ensure consistent outputs. 
%%    \item SciBERT \cite{beltagy2019scibert}, which implements an encoder transformer trained on scientific documents. We used the named entity recognition implementation from the Hugging Face library developed by AllenAI\footnote{\url{https://huggingface.co/allenai/scibert_scivocab_cased}}. 
%%    \item SpaCy \cite{spacyner}, which implements incremental parsing with Bloom embeddings and residual convolutional neural networks. Entity recognition is performed by invoking the entity recognition module from the Hugging Face library\footnote{\url{https://huggingface.co/spacy/en_core_web_sm}}.
%%\end{itemize}
%%Although the NER performance of them has been evaluated in previous papers, they used different datasets. 

%%\begin{table}
%%\caption{Performance comparison on scientific entity extraction across three models. }\label{table:ner}
%%\begin{tabular}{llll}
%%\toprule
%%Model     &  Precision & Recall & F1\\
%%\midrule
%%GPT-4     &0.60 & 0.79  &0.66 \\
%%SciBERT & 0.04&0.08&0.05\\
%%SpaCy &0.08&0.06&0.07\\   
%%\bottomrule
%%\end{tabular}
%%\end{table}

\begin{figure}
    \centering
    \includegraphics[width=0.5\textwidth]{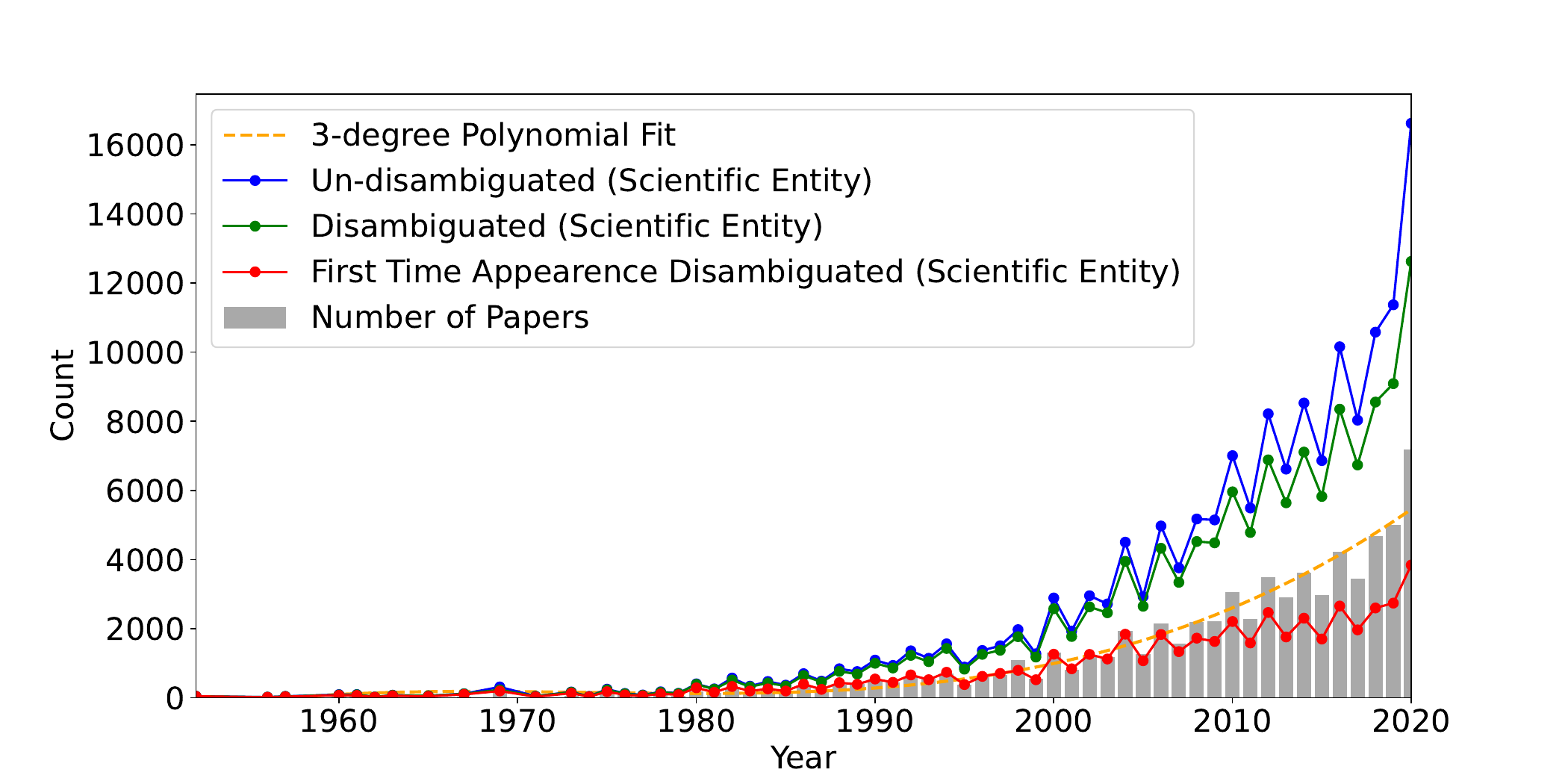}
    \caption{The number of papers, scientific entities (undisambiguated), and disambiguated entities in the ACL Corpus.}
    \label{fig:paperentityyear}
\end{figure}

To compare the performance of these models, we built a small benchmark dataset by manually annotating 200 titles randomly selected from all ACL Anthology papers, following the annotation guidelines in Wu et al. \cite{wu2020jcdl}. 
%%The precision, recall, and F1-score for each model is shown in Table~\ref{table:ner}. 
The F1-scores achieved by GPT-4, SciBERT, and SpaCy are $0.66$, $0.05$, and $0.07$, respectively, indicating that GPT-4 outperforms the other two models, so we adopt GPT-4 for recognizing sciientific entities of all paper titles. The numbers of papers and scientific entities recognized are shown in Figure~\ref{fig:paperentityyear}. The average number of scientific entities per title is about 3. 

\subsection{Scientific Entity Disambiguation}
Using the method described in Section~\ref{sec:ner}, the threshold was calibrated based on the classification performance evaluated against 180 manually labeled entity pairs. Entity pairs extracted from titles within a quota were automatically labeled as ``similar" or ``not similar" based on the calculated similarity scores and a threshold of 0.5. 
\subsection{Document Frequency Curve Fitting}
\begin{figure}
    \centering
    \includegraphics[width=\linewidth]{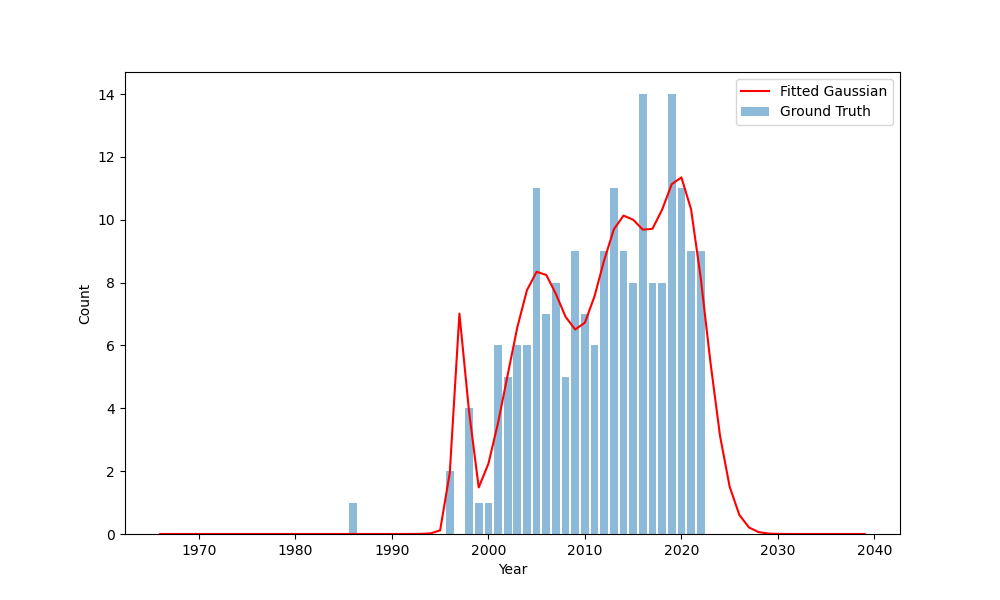}
    \caption{The document frequency chart (blue) of an entity named \emph{machine learning} and a fitting with {4} Gaussian profiles.
    \label{fig:visualzation of fitting}}
\end{figure}

For each scientific entity $e$, we count the number of paper titles that contain $e$ at Year $t$ and plot the document frequency chart. An example is shown in Figure~\ref{fig:visualzation of fitting}. 

The document frequency chart for each scientific entity is fit using a composite of Gaussian profiles each having three parameters, peak, mean, and dispersion. Our fitting program employs a dynamic tuning approach, where the number of peaks is inferred from the data using an algorithm based on a comparison of neighboring values\footnote{\url{https://docs.scipy.org/doc/scipy/reference/generated/scipy.signal.find_peaks.html}}. 
%%, calculated by if the word frequency is below 500, then epochs are 5000, if it is above 500, then use frequency * 10 to be the epoches, if the loss is lower than 0.4, then early stop the fitting} 
%%The Gaussian parameters are initialized based on detected peaks and statistical analysis of the observed data. Specifically, t
The center of each Gaussian is initialized at the detected peak position; the amplitude is initialized randomly within a defined range, and the width of each Gaussian is initialized based on the year range divided by the number of peaks. The fitting process iteratively updates the parameters using gradient descent. We used ADAM as the optimizer \cite{diederik2014adam} and the mean squared error (MSE) as the loss function. To prevent overfitting, a regularization term is added to the loss function, which penalizes excessively large or narrow peaks. The number of epochs and parameters was automatically adjusted by the fitting algorithm.

\subsection{Calculating FICE}
After fitting the document frequency chart for a scientific entity $e$, the starting point for $t_s$ is determined as the year when $e$ first appeared and $t_e$ {is determined when the predicted document frequency is less than 1}, which may be beyond the time span of the observable period. We calculated the lifetime ratio for each $e$ in a title $d$ using Eq.~(\ref{eq:r}) and the informativeness weight using Eq.~(\ref{eq:w}). The FICE for a given quota $Q$ is calculated using Eq.~(\ref{eq:cce}).

\subsection{Cumulative Citation Count $C_5$}
%%As mentioned before, raw citation count is in general not a reliable factor due to its measure for being noisy, biased, and domain-dependent. However, it was argued that short-term citations can be considered as currency on the research front and long-term citations can contribute to the codification of knowledge claims into bodies of knowledge \cite{leydesdorff2016citations}. Determining the exact boundary between short-term and long-term is non-trivial. Here we adopt $C_5 (y)$, which is the 5-year average cumulative citation count of papers within a quota as an estimate of the quota's average impact over the short- and long-term. Our study focuses on the ACL Anthology, so the domain dependency is relatively weak. In the future, we will delineate short-term and long-term citations using the Multi-RPYS method \cite{leydesdorff2016citations}. 
We obtain citations for each paper each year from the Semantic Scholarly Graph API \cite{s2graphapi}. Therefore, the average 5-year cumulative citation count in 2015 is $C_5 (2015)=\sum_{y=2015}^{2019}c(y)$, in which $c(y)$ is the citation received by a paper in Year~$y$. 

\begin{figure}
    \centering
    \includegraphics[width=0.47\textwidth]{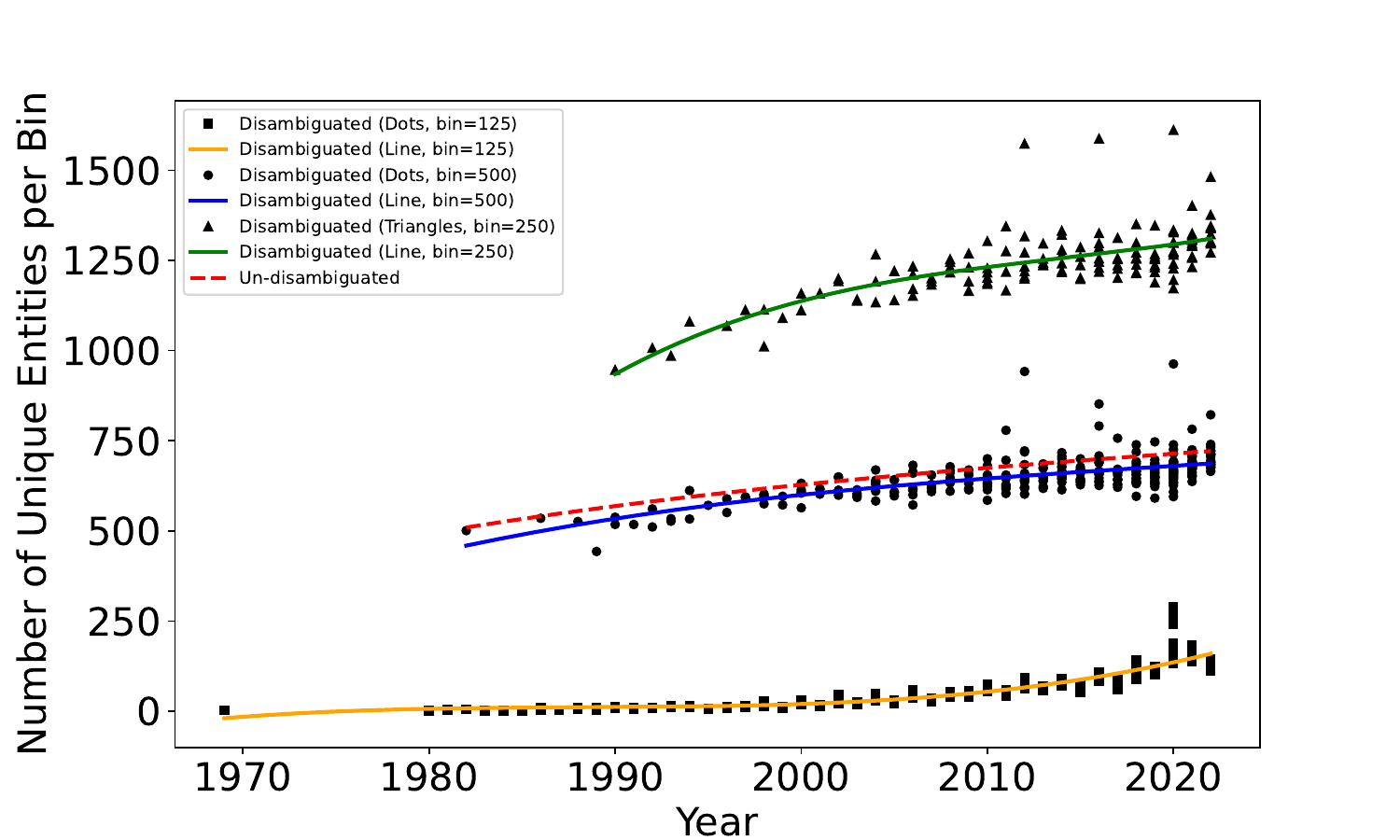}
    \caption{The FICE calculated using disambiguated scientific entities (black dots) with $|Q|=125, 250, 500$. The red and blue curves are the polynomial fittings of disambiguated and undisambiguated entities, respectively. For each year, we only plot data points that represent full quotas of papers. }
    \label{fig:enter-label}
\end{figure}

\section{Results}
\subsection{Growth of Scientific Entity Diversity}
To distinguish from the original cognitive extent, we define entity-based cognitive extent as the unique number of scientific entities extracted from paper titles within a quota. Similar to lexical diversity, entity-based cognitive extent can be seen as a measure of the scientific entity diversity. By plotting entity-based cognitive extent over time, we found that it gradually increases with time, which is consistent with the trend of the original cognitive extent based on paper titles in Astronomy, Physics, and Biomedical \cite{milojevic2015quantifying}.  We compared the trends calculated using disambiguated and undisambiguated scientific entities (Figure~\ref{fig:ccec5undisamb}) and found that undisambiguated entities linearly shifted the curve up but did not significantly change the growth rate. 
%%This is consistent with the observation made by Milojevic using the original cognitive extent extracted from paper titles in Astronomy, Physics, and Biomedicine \cite{milojevic2015quantifying}, 

To investigate how the growth rate of entity-based cognitive extent increases with time, we fit the disambiguated entity-based data points using a linear function from 1980 to 2000 and another linear function from 2000 to 2020, respectively. The slopes obtained for various quotas are tabulated in Table~\ref{tab:slope}. The results indicate that the entity-based cognitive extent increases at a slower rate. 

The entity-based cognitive extent vs. year relations for three quota sizes are illustrated in Figure~\ref{fig:ccec5undisamb}. Similar to the original cognitive extent, calculating the entity-based cognitive extent within a quota is important to generate a consistent measure of the metric. Our quota is different from the ones used in Milojevi\'{c} in two aspects. First, instead of a fixed number of phrases, we use a fixed number of titles. This is because the informativity weight is normalized within scientific entities of a particular title before weights of multiple titles are aggregated. Second, the quota sizes used in our study are smaller than the quota used in the original cognitive extent, which is 3000 -- 10000. For example, $|Q|=500$ converts to about 1500 scientific entities. Low quota data points may suffer a saturation, meaning that cognitive extent value increases with quota size for a given time point, which is seen in  Figure~\ref{fig:ccec5undisamb}. To fix the problem, a linear correction factor can be applied to ``lift'' data points in low quota to be aligned with unsaturated data points. The exact correction factor may be domain-dependent and will be investigated with a larger corpus in future studies. 

\begin{table}
    \centering
    \caption{The linear fit slopes of entity-based cognitive extents.}
    \begin{tabular}{c|ccc}
    \toprule
      Year Range   & $\mathbf{|Q|=125}$ &$\mathbf{|Q|=250}$ &$\mathbf{|Q|=500}$\\
      \midrule
       1980--2000  & 1.19 & 7.21 & 15.88  \\
       \midrule
       2000--2020 & 6.80 & 3.09 & 6.10\\
       \bottomrule
    \end{tabular}
    
    \label{tab:slope}
\end{table}\

%%\begin{figure}
%%    \centering
%%    \includegraphics[width=0.47\textwidth]{figs/area-bin250-0.7932-1.716e-07.png}
%%    \includegraphics[width=0.47\textwidth]{figs/area-bin500-0.8095-0.00014.png}
%%    \caption{CCE vs. $C_5$ for a quota size of 250 (top) or 500 (bottom). }
%%    \label{fig:ccec5}
%%\end{figure}

\subsection{Correlation Between FICE with $C_5$}
We found that FICE exhibits a strong positive correlation with $\log{C_5}$ as shown in Figure~\ref{fig:ccec5undisamb}. The plots were made by first ranking papers by $C_5(2015)$ in ascending order and binning the sequence by a given quota. For each data point, the $x$-coordinate is calculated as the logarithmic value of the average of $C_5 (2015)$ of papers in a quota, and the $y$-coordinate is {the arithmetic average FICE of papers in the same quota. The error bars are calculated as the standard deviation assuming a Gaussian distribution. The Spearman correlation coefficients between FICE and $\log(C_5)$ for three quota are shown in Table~\ref{tab:ablation}.
%%is $\rho_{125} = 0.783$ ($p < 0.001$), $\rho_{250} = 0.756$ ($p = 0.007$), and $\rho_{500} = 0.717$ ($p = 0.109$).
To test the influence of disambiguation on the correlation. We calculated FICE without entity disambiguation and obtained similar Spearman correlation coefficients.

\begin{figure}
    \centering
    \includegraphics[width=0.47\textwidth]{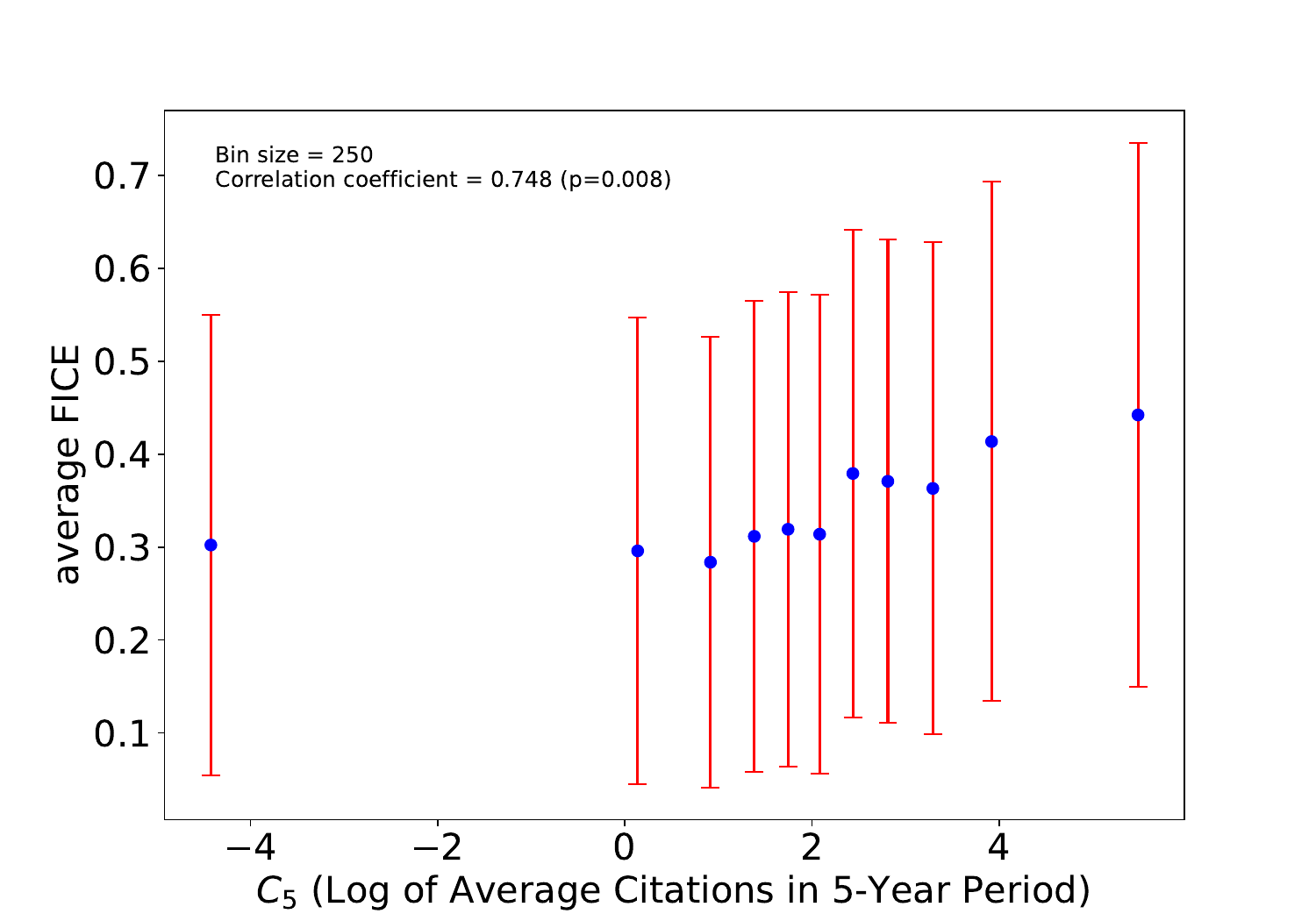}
    \caption{Average FICE calculated using undisambiguated entities per quota vs. the $\log{C_5}$. Paper titles are grouped into a bin size of $250$.}
    \label{fig:ccec5undisamb}
\end{figure}

%%We clarify that freshness is not equivalent to novelty, which measures the \emph{quality} of being new, original, or unusual. Being fresh means that the entity is at the early stage of its lifetime, but the entity may have been proposed for many years and thus not been novel anymore. 
Note that Figure~\ref{fig:ccec5undisamb} reveals a \emph{collective} instead of an individual correlation because FICE represents the weighted cognitive extent in a quota. This correlation does not apply to individual papers because of the small number of scientific entities in a title. Therefore, the correlation does not imply that one could increase the citation impact by simply using never-existing entity names. If the paper lacks true novelty and significant contributions, newly introduced entity names are unlikely to be adopted in subsequent research, rendering them ``transient'' with a short lifetime and a minimal contribution to FICE.

%%\subsection{Baseline Comparison}
We compare FICE with three simplified versions and demonstrate the contribution of the lifetime ratio and informativity weight in the correlation above. These simplified versions are (1) \textbf{Dichotomous Entity-based Cognitive Extent}, calculated by adding the number of disambiguated unique scientific entities in a quota. (2) \textbf{Weight Only}, calculated by simply summing up the normalized weights (Eq.(\ref{eq:w})) of all scientific entities. %%without multiplying by the lifetime ratio factor. 
    (3) \textbf{Lifetime Ratio Only}, calculated by summing up the unweighted lifetime ratios $1-r(e,t)$ of all scientific entities. 
Table~\ref{tab:ablation} indicates that FICE exhibits the strongest correlation against all baseline models. Both the lifetime ratio and the informativity weight contribute to this strong correlation. The quota size will influence the correlation coefficients. In particular, all simplified versions exhibit a strong correlation with $|Q|=500$. The lifetime ratio consistently exhibits a strong correlation with $\log{C_5}$.

\begin{table}[h]
\centering
\caption{Speareman correlation coefficients and $p$-values between FICE and three simplified versions:  Dicho (Dichotomous), Weight (Weight only), and L. Ratio (Lifetime Ratio only). }
\label{tab:ablation}
\begin{tabular}{lrrr}
\toprule
{\bf Method} & $\mathbf{|Q|=125}$ & $\mathbf{|Q|=250}$ & $\mathbf{|Q|=500}$ \\
\midrule
Dicho & $-0.206 (0.371)$ & $-0.333 (0.317)$ & $-0.703 (0.119)$ \\
Weight & $0.228 (0.320)$ & $0.334 (0.316)$ & $0.715 (0.110)$ \\
L. Ratio & $0.603 (0.004)$ & $0.705 (0.015)$ & $0.744 (0.090)$ \\
\midrule
{\bfseries FICE} & {\bfseries $0.766 (<0.001)$} & {\bfseries $0.748 (0.008)$} & {\bfseries $0.717 (0.109)$} \\
\bottomrule
\end{tabular}
\end{table}

\section{Conclusion}
We proposed FICE, which extends the original cognitive extent. FICE is calculated based on the lifetime ratio and informativity of scientific entities extracted from paper titles within a quota. Using ACL Anthology, we found the number of unique scientific entities per quota increased with time, consistent with previous observations in other disciplines. FICE exhibits a strong positive correlation with the average 5-year cumulative citation count, which may be used for predicting collective citations for trending topics.

\section{Acknowledgments}
The paper is presented at the second Workshop on “Innovation Measurement for Scientific Communication (IMSC) in the Era of Big Data” at 2024 ACM/IEEE Joint Conference on Digital Libraries (JCDL).

\bibliography{main}

\begin{thebibliography}{24}
\expandafter\ifx\csname natexlab\endcsname\relax\def\natexlab#1{#1}\fi
\providecommand{\url}[1]{\texttt{#1}}
\providecommand{\href}[2]{#2}
\providecommand{\path}[1]{#1}
\providecommand{\DOIprefix}{doi:}
\providecommand{\ArXivprefix}{arXiv:}
\providecommand{\URLprefix}{URL: }
\providecommand{\Pubmedprefix}{pmid:}
\providecommand{\doi}[1]{\href{http://dx.doi.org/#1}{\path{#1}}}
\providecommand{\Pubmed}[1]{\href{pmid:#1}{\path{#1}}}
\providecommand{\bibinfo}[2]{#2}
\ifx\xfnm\relax \def\xfnm[#1]{\unskip,\space#1}\fi
%Type = Article
\bibitem[{Milojevi{\'c}(2015)}]{milojevic2015quantifying}
\bibinfo{author}{S.~Milojevi{\'c}},
\newblock \bibinfo{title}{Quantifying the cognitive extent of science},
\newblock \bibinfo{journal}{Journal of Informetrics} \bibinfo{volume}{9} (\bibinfo{year}{2015}) \bibinfo{pages}{962--973}.
%Type = Article
\bibitem[{Bérubé et~al.(2018)Bérubé, Sainte-Marie, Mongeon, and Larivière}]{berube2018}
\bibinfo{author}{N.~Bérubé}, \bibinfo{author}{M.~Sainte-Marie}, \bibinfo{author}{P.~Mongeon}, \bibinfo{author}{V.~Larivière},
\newblock \bibinfo{title}{Words by the tail: Assessing lexical diversity in scholarly titles using frequency-rank distribution tail fits},
\newblock \bibinfo{journal}{PLOS ONE} \bibinfo{volume}{13} (\bibinfo{year}{2018}) \bibinfo{pages}{1--31}. \URLprefix \url{https://doi.org/10.1371/journal.pone.0197775}. \DOIprefix\doi{10.1371/journal.pone.0197775}.
%Type = Article
\bibitem[{Jarvis and Daller(2013)}]{jarvis2013defining}
\bibinfo{author}{S.~Jarvis}, \bibinfo{author}{M.~Daller},
\newblock \bibinfo{title}{Defining and measuring lexical diversity},
\newblock \bibinfo{journal}{Vocabulary knowledge: Human ratings and automated measures. Amsterdam, The Netherlands}  (\bibinfo{year}{2013}).
%Type = Book
\bibitem[{Malvern et~al.(2004)Malvern, Richards, Chipere, and Dur{\'a}n}]{malvern2004lexical}
\bibinfo{author}{D.~Malvern}, \bibinfo{author}{B.~Richards}, \bibinfo{author}{N.~Chipere}, \bibinfo{author}{P.~Dur{\'a}n}, \bibinfo{title}{Lexical diversity and language development}, \bibinfo{publisher}{Springer}, \bibinfo{year}{2004}.
%Type = Article
\bibitem[{Milojević(2017)}]{milojevic2017length}
\bibinfo{author}{S.~Milojević},
\newblock \bibinfo{title}{The length and semantic structure of article titles—evolving disciplinary practices and correlations with impact},
\newblock \bibinfo{journal}{Frontiers in Research Metrics and Analytics} \bibinfo{volume}{2} (\bibinfo{year}{2017}). \URLprefix \url{https://www.frontiersin.org/journals/research-metrics-and-analytics/articles/10.3389/frma.2017.00002}. \DOIprefix\doi{10.3389/frma.2017.00002}.
%Type = Article
\bibitem[{Eykens et~al.(2023)Eykens, Guns, Engels, and Vandermoere}]{eykens2023cognitive}
\bibinfo{author}{J.~Eykens}, \bibinfo{author}{R.~Guns}, \bibinfo{author}{T.~C. Engels}, \bibinfo{author}{F.~Vandermoere},
\newblock \bibinfo{title}{Cognitive and interdisciplinary mobility in the social sciences and humanities: Traces of increased boundary crossing},
\newblock \bibinfo{journal}{Journal of Information Science}  (\bibinfo{year}{2023}) \bibinfo{pages}{01655515231171086}. \URLprefix \url{https://doi.org/10.1177/01655515231171086}. \DOIprefix\doi{10.1177/01655515231171086}.
%Type = Article
\bibitem[{Waltman(2016)}]{waltman2016review}
\bibinfo{author}{L.~Waltman},
\newblock \bibinfo{title}{A review of the literature on citation impact indicators},
\newblock \bibinfo{journal}{Journal of Informetrics} \bibinfo{volume}{10} (\bibinfo{year}{2016}) \bibinfo{pages}{365--391}. \URLprefix \url{https://www.sciencedirect.com/science/article/pii/S1751157715300900}. \DOIprefix\doi{https://doi.org/10.1016/j.joi.2016.02.007}.
%Type = Article
\bibitem[{Albarr{\'a}n et~al.(2011)Albarr{\'a}n, Crespo, Ortu{\~n}o, and Ruiz-Castillo}]{albarran2011skewness}
\bibinfo{author}{P.~Albarr{\'a}n}, \bibinfo{author}{J.~A. Crespo}, \bibinfo{author}{I.~Ortu{\~n}o}, \bibinfo{author}{J.~Ruiz-Castillo},
\newblock \bibinfo{title}{The skewness of science in 219 sub-fields and a number of aggregates},
\newblock \bibinfo{journal}{Scientometrics} \bibinfo{volume}{88} (\bibinfo{year}{2011}) \bibinfo{pages}{385--397}.
%Type = Article
\bibitem[{Leydesdorff et~al.(2016)Leydesdorff, Bornmann, Comins, and Milojevi{\'c}}]{leydesdorff2016citations}
\bibinfo{author}{L.~Leydesdorff}, \bibinfo{author}{L.~Bornmann}, \bibinfo{author}{J.~A. Comins}, \bibinfo{author}{S.~Milojevi{\'c}},
\newblock \bibinfo{title}{Citations: Indicators of quality? the impact fallacy},
\newblock \bibinfo{journal}{Frontiers in Research Metrics and Analytics} \bibinfo{volume}{1} (\bibinfo{year}{2016}). \URLprefix \url{https://www.frontiersin.org/journals/research-metrics-and-analytics/articles/10.3389/frma.2016.00001}. \DOIprefix\doi{10.3389/frma.2016.00001}.
%Type = Inproceedings
\bibitem[{Wu et~al.(2020)Wu, Ul~Hoque, Reiske, Weigle, Bradshaw, Gaff, Li, and Kwan}]{wu2020comparative}
\bibinfo{author}{J.~Wu}, \bibinfo{author}{M.~R. Ul~Hoque}, \bibinfo{author}{G.~W. Reiske}, \bibinfo{author}{M.~C. Weigle}, \bibinfo{author}{B.~T. Bradshaw}, \bibinfo{author}{H.~D. Gaff}, \bibinfo{author}{J.~Li}, \bibinfo{author}{C.~Kwan},
\newblock \bibinfo{title}{A comparative study of sequence tagging methods for domain knowledge entity recognition in biomedical papers},
\newblock in: \bibinfo{booktitle}{Proceedings of the ACM/IEEE Joint Conference on Digital Libraries in 2020}, \bibinfo{year}{2020}, pp. \bibinfo{pages}{397--400}.
%Type = Inproceedings
\bibitem[{Hong et~al.(2020)Hong, Tchoua, Chard, and Foster}]{hong2020sciner}
\bibinfo{author}{Z.~Hong}, \bibinfo{author}{R.~Tchoua}, \bibinfo{author}{K.~Chard}, \bibinfo{author}{I.~Foster},
\newblock \bibinfo{title}{Sciner: Extracting named entities from scientific literature},
\newblock in: \bibinfo{editor}{V.~V. Krzhizhanovskaya}, \bibinfo{editor}{G.~Z{\'a}vodszky}, \bibinfo{editor}{M.~H. Lees}, \bibinfo{editor}{J.~J. Dongarra}, \bibinfo{editor}{P.~M.~A. Sloot}, \bibinfo{editor}{S.~Brissos}, \bibinfo{editor}{J.~Teixeira} (Eds.), \bibinfo{booktitle}{Computational Science -- ICCS 2020}, \bibinfo{publisher}{Springer International Publishing}, \bibinfo{address}{Cham}, \bibinfo{year}{2020}, pp. \bibinfo{pages}{308--321}.
%Type = Article
\bibitem[{Li et~al.(2020)Li, Sun, Han, and Li}]{li2020nersurvey}
\bibinfo{author}{J.~Li}, \bibinfo{author}{A.~Sun}, \bibinfo{author}{J.~Han}, \bibinfo{author}{C.~Li},
\newblock \bibinfo{title}{A survey on deep learning for named entity recognition},
\newblock \bibinfo{journal}{IEEE transactions on knowledge and data engineering} \bibinfo{volume}{34} (\bibinfo{year}{2020}) \bibinfo{pages}{50--70}.
%Type = Article
\bibitem[{Wang et~al.(2023)Wang, Sun, Li, Ouyang, Wu, Zhang, Li, and Wang}]{wang2023gptner}
\bibinfo{author}{S.~Wang}, \bibinfo{author}{X.~Sun}, \bibinfo{author}{X.~Li}, \bibinfo{author}{R.~Ouyang}, \bibinfo{author}{F.~Wu}, \bibinfo{author}{T.~Zhang}, \bibinfo{author}{J.~Li}, \bibinfo{author}{G.~Wang},
\newblock \bibinfo{title}{Gpt-ner: Named entity recognition via large language models},
\newblock \bibinfo{journal}{arXiv preprint arXiv:2304.10428}  (\bibinfo{year}{2023}).
%Type = Inproceedings
\bibitem[{Sousa et~al.(2023)Sousa, Guimarães, Jorge, and Campos}]{sousa2023ner}
\bibinfo{author}{H.~Sousa}, \bibinfo{author}{N.~Guimarães}, \bibinfo{author}{A.~Jorge}, \bibinfo{author}{R.~Campos},
\newblock \bibinfo{title}{Gpt struct me: Probing gpt models on narrative entity extraction},
\newblock in: \bibinfo{booktitle}{2023 IEEE/WIC International Conference on Web Intelligence and Intelligent Agent Technology (WI-IAT)}, \bibinfo{year}{2023}, pp. \bibinfo{pages}{383--387}. \DOIprefix\doi{10.1109/WI-IAT59888.2023.00063}.
%Type = Article
\bibitem[{Li et~al.(2020)Li, Li, Suhara, Doan, and Tan}]{li2020deep}
\bibinfo{author}{Y.~Li}, \bibinfo{author}{J.~Li}, \bibinfo{author}{Y.~Suhara}, \bibinfo{author}{A.~Doan}, \bibinfo{author}{W.-C. Tan},
\newblock \bibinfo{title}{Deep entity matching with pre-trained language models},
\newblock \bibinfo{journal}{arXiv preprint arXiv:2004.00584}  (\bibinfo{year}{2020}).
%Type = Book
\bibitem[{De~Beaugrande and Dressler(1981)}]{de1981introduction}
\bibinfo{author}{R.-A. De~Beaugrande}, \bibinfo{author}{W.~U. Dressler}, \bibinfo{title}{Introduction to text linguistics}, volume~\bibinfo{volume}{1}, \bibinfo{publisher}{longman London}, \bibinfo{year}{1981}.
%Type = Article
\bibitem[{Tocatlian(1970)}]{tocatlian1970titles}
\bibinfo{author}{J.~J. Tocatlian},
\newblock \bibinfo{title}{Are titles of chemical papers becoming more informative?},
\newblock \bibinfo{journal}{Journal of the American Society for Information Science} \bibinfo{volume}{21} (\bibinfo{year}{1970}) \bibinfo{pages}{345--350}.
%Type = Article
\bibitem[{Radev et~al.(2013)Radev, Muthukrishnan, Qazvinian, and Abu-Jbara}]{radev2013acl}
\bibinfo{author}{D.~R. Radev}, \bibinfo{author}{P.~Muthukrishnan}, \bibinfo{author}{V.~Qazvinian}, \bibinfo{author}{A.~Abu-Jbara},
\newblock \bibinfo{title}{The acl anthology network corpus},
\newblock \bibinfo{journal}{Language Resources and Evaluation} \bibinfo{volume}{47} (\bibinfo{year}{2013}) \bibinfo{pages}{919--944}.
%Type = Article
\bibitem[{Brown(2020)}]{brown2020language}
\bibinfo{author}{T.~B. Brown},
\newblock \bibinfo{title}{Language models are few-shot learners},
\newblock \bibinfo{journal}{arXiv preprint arXiv:2005.14165}  (\bibinfo{year}{2020}).
%Type = Inproceedings
\bibitem[{Beltagy et~al.(2019)Beltagy, Lo, and Cohan}]{beltagy2019scibert}
\bibinfo{author}{I.~Beltagy}, \bibinfo{author}{K.~Lo}, \bibinfo{author}{A.~Cohan},
\newblock \bibinfo{title}{{S}ci{BERT}: A pretrained language model for scientific text},
\newblock in: \bibinfo{booktitle}{Proceedings of the 2019 Conference on Empirical Methods in Natural Language Processing and the 9th International Joint Conference on Natural Language Processing (EMNLP-IJCNLP)}, \bibinfo{publisher}{Association for Computational Linguistics}, \bibinfo{address}{Hong Kong, China}, \bibinfo{year}{2019}, pp. \bibinfo{pages}{3615--3620}. \URLprefix \url{https://www.aclweb.org/anthology/D19-1371}. \DOIprefix\doi{10.18653/v1/D19-1371}.
%Type = Misc
\bibitem[{{Matthew Honnibal}(2024)}]{spacyner}
\bibinfo{author}{{Matthew Honnibal}}, \bibinfo{title}{{spaCy's NER Model}}, \bibinfo{year}{2024}.
%Type = Inproceedings
\bibitem[{Wu et~al.(2020)Wu, Ul~Hoque, Reiske, Weigle, Bradshaw, Gaff, Li, and Kwan}]{wu2020jcdl}
\bibinfo{author}{J.~Wu}, \bibinfo{author}{M.~R. Ul~Hoque}, \bibinfo{author}{G.~W. Reiske}, \bibinfo{author}{M.~C. Weigle}, \bibinfo{author}{B.~T. Bradshaw}, \bibinfo{author}{H.~D. Gaff}, \bibinfo{author}{J.~Li}, \bibinfo{author}{C.~Kwan},
\newblock \bibinfo{title}{A comparative study of sequence tagging methods for domain knowledge entity recognition in biomedical papers},
\newblock in: \bibinfo{booktitle}{Proceedings of the ACM/IEEE Joint Conference on Digital Libraries in 2020}, JCDL '20, \bibinfo{publisher}{Association for Computing Machinery}, \bibinfo{address}{New York, NY, USA}, \bibinfo{year}{2020}, p. \bibinfo{pages}{397–400}. \URLprefix \url{https://doi.org/10.1145/3383583.3398602}. \DOIprefix\doi{10.1145/3383583.3398602}.
%Type = Inproceedings
\bibitem[{Kingma and Ba(2015)}]{diederik2014adam}
\bibinfo{author}{D.~P. Kingma}, \bibinfo{author}{J.~Ba},
\newblock \bibinfo{title}{Adam: {A} method for stochastic optimization},
\newblock in: \bibinfo{editor}{Y.~Bengio}, \bibinfo{editor}{Y.~LeCun} (Eds.), \bibinfo{booktitle}{3rd International Conference on Learning Representations, {ICLR} 2015, San Diego, CA, USA, May 7-9, 2015, Conference Track Proceedings}, \bibinfo{year}{2015}. \URLprefix \url{http://arxiv.org/abs/1412.6980}.
%Type = Misc
\bibitem[{{AllenAI}(2024)}]{s2graphapi}
\bibinfo{author}{{AllenAI}}, \bibinfo{title}{{Semantic Scholar API}}, \bibinfo{howpublished}{\url{https://www.semanticscholar.org/product/api}}, \bibinfo{year}{2024}.

\end{thebibliography}
\bibliographystyle{unsrt}

%%
%% If your work has an appendix, this is the place to put it.
\appendix

%%\section{Online Resources}

\end{document}